\documentclass[aps,prb,preprint,superscriptaddress,showpacs]{revtex4-1}
\usepackage{graphicx}% Include figure files
\usepackage{caption}		%change captions
\usepackage{subcaption}
\usepackage{bm}% bold math
\usepackage{amsmath}% \text{}

\begin{document}

%\preprint{}

\title{Spin/orbit moment imbalance in the near-zero moment ferromagnetic semiconductor SmN}


\author{Eva-Maria Anton}
\email[]{eva.anton@vuw.ac.nz}
%\thanks{}

\author{B. J. Ruck}

\affiliation{MacDiarmid Institute of Advanced Materials and Nanotechnology, School of Chemical and Physical Sciences, Victoria University, PO Box 600, Wellington, New Zealand}

\author{C. Meyer}
\affiliation{Institut N\'{e}el, Centre National de la Recherche Scientifique and Universit\'{e} Joseph Fourier, Bo\^{i}te Postale 166, F-38042 Grenoble CEDEX, France}

\author{F. Natali}
\author{Harry Warring}

\affiliation{MacDiarmid Institute of Advanced Materials and Nanotechnology, School of Chemical and Physical Sciences, Victoria University, PO Box 600, Wellington, New Zealand}

\author{Fabrice Wilhelm}
\author{A. Rogalev}
\affiliation{European Synchrotron Radiation Facility, Bo\^{i}te Postale 220, F-38043 Grenoble CEDEX, France}

\author{V. N. Antonov}
\affiliation{Institute of Metal Physics, National Academy of Sciences of Ukraine, 03142 Kiev, Ukraine}

\author{H. J. Trodahl}

\affiliation{MacDiarmid Institute of Advanced Materials and Nanotechnology, School of Chemical and Physical Sciences, Victoria University, PO Box 600, Wellington, New Zealand}


\date{\today}

\begin{abstract}
SmN is ferromagnetic below 27 K, and its net magnetic moment of 0.03 Bohr magnetons per formula unit is 
one of the smallest magnetisations found in any ferromagnetic material. The near-zero moment is a result 
of the nearly equal and opposing spin and orbital moments in the $^6$H$_{5/2}$ ground state of the Sm$^{3+}$ ion, 
which leads finally to a nearly complete cancellation for an ion in the SmN ferromagnetic state. Here we 
explore the spin alignment in this compound with X-ray magnetic circular dichroism at the Sm L$_{2,3}$ edges. 
The spectral shapes are in qualitative agreement with computed spectra based on an LSDA+$U$ (local spin density approximation
with Hubbard-U corrections) band 
structure, though there remain differences in detail which we associate with the anomalous branching 
ratio in rare-earth L edges. The sign of the spectra determine that in a magnetic field the Sm 4$f$ spin 
moment aligns antiparallel to the field; the very small residual moment in ferromagnetic SmN aligns with 
the 4$f$ orbital moment and antiparallel to the spin moment. Further measurements on very thin (1.5 nm) 
SmN layers embedded in GdN show the opposite alignment due to a strong Gd-Sm exchange, suggesting that 
the SmN moment might be further reduced by about 0.5 \% Gd substitution. 
\end{abstract}

%\keywords{}

\maketitle


\section{\label{sec:introduction}Introduction}

The rare earths, across which the 4$f$ level fills, provide an enormous range of interesting and potentially exploitable spin/orbit physics. In elemental form they are metallic, but they are also prominent as the sites of magnetic order in a variety of insulating compounds, where furthermore transitions among internal spin-orbit configurations in their 4$f$ shells allow for a rich optical activity. Recently there has been a renewed interest in these elements in their simple NaCl structured mononitrides, with theoretical suggestions of ferromagnetic (FM) semiconductors and half metals.~\cite{PhysRevB.75.045114,PhysRevB.69.045115} Experimental verification of the predictions is lacking for most of the series, though recent studies have identified several as intrinsic FM semiconductors. The most thoroughly studied is GdN, lying at the centre of the series. The half-filled 4$f$ shell on the Gd$^{3+}$ ions have (L,S) of (0,7/2) ($^{8}$S$_{7/2}$) and a magnetic moment of 7~$\mu_B$. GdN is FM below a Curie temperature (T$_c$) most often found to be near 65-70 K, and with 32 aligned Gd ions per nm$^3$ it has one of the largest magnetizations known.~\cite{PhysRevB.72.014427,PhysRevB.73.235335} The next lightest rare-earth, Eu$^{3+}$, has a J = 0 ground state ($^7$F$_0$), which then prevents any magnetic order, though its propensity for mixed valence leads to unusual magnetic behaviour in EuN.~\cite{PhysRevB.83.174404} Sm$^{3+}$, at one lighter again, has a ground state of $^6$H$_{5/2}$ and a 4$f$ magnetic moment of $M\approx {\mu}_{B}(L+2S) \approx 0$. In the free ion the moment is 0.8~$\mu_B$, but it is commonly smaller in the crystal fields of solids. 

The earliest magnetic investigations of SmN, performed forty years ago, found a very small moment below a transition near 25 K. It was then assigned as antiferromagnetic based on a vanishingly small net moment in the ordered phase.~\cite{hulliger1978,vogt1993} Subsequently a neutron scattering study suggested instead a FM phase with near cancellation between the spin and orbital moments.~\cite{Moon1979265} A FM state with a 27 K T$_c$ has recently been confirmed by magnetisation data on a thin film.~\cite{PhysRevB.78.174406} That study showed that in the crystal field of SmN the paramagnetic (PM) moment is 0.45~$\mu_B$, smaller than the free ion moment. In the FM state it drops to the remarkably small value of ${{\sim}}0.03~\mu_B$. The moment then couples only weakly to an external field, so that the coercive field rises above the 6~T limit that was available in the reported SQUID study. 

In the past decade there have been a number of reports devoted to metallic Sm compounds that similarly have near zero net moments in their FM phases.~\cite{PhysRevB.59.11445,He2011985} In these, as in SmN, the 4$f$ spins order ferromagnetically but with the resulting spin moment largely cancelled by the 4$f$ orbital contribution. In addition to the intrinsic fundamental interest in such moment-free FM materials, these show promise as fringe-field free sources of spin-polarised electrons for injection into conventional semiconductors, or for super-hard FM layers to pin the moments of softer magnetic layers. Among these small-moment FM materials, SmAl$_2$ has had special attention. It has a modest net moment in the FM phase of 0.26~$\mu_B$ per formula unit, oriented anti-parallel with the 4$f$ spin moment and thus parallel with the 4$f$ orbital moment. Its net moment is larger by a factor of about 5 in comparison with, for example, SmZn or SmCd, but in contrast these have net moments that are parallel to the 4$f$ spin moment.~\cite{PhysRevB.59.11445} It is that difference that has encouraged the attention paid to SmAl$_2$, for the ease with which the spin moment can be increased, with the introduction of substitutional Gd ions, to fully compensate the orbital moment.~\cite{PhysRevLett.87.127202,PhysRevB.70.134418,Avisou2008,PhysRevB.82.174421,maji2011,Adachi1999} Since in any case the net 4$f$ moment on Sm$^{3+}$ shows a temperature dependence from admixture with the J=7/2 excited state at 1500 K, exact compensation finally occurs at only one crossover temperature. It will be shown below that FM SmN shares with SmAl$_2$ the character that the net moment lies antiparallel to the 4$f$ spin moment. 

Interestingly an LSDA+$U$ (local spin density approximation with Hubbard-U corrections) calculation also suggests a moment of 0.03~$\mu_B$ per Sm ion in the FM phase of SmN.~\cite{PhysRevB.75.045114} That remarkably small moment results from not only a near cancellation between 4$f$ spin and orbital contributions, each of about 5~$\mu_B$, but also has contributions of order 0.1~$\mu_B$ from the Sm 5$d$ and N 2$p$ shells. The agreement is surprising, in view of the $<1~\%$ computational accuracy that would be required. The calculation gave a net magnetisation coincident with the 4$f$ spin moment direction, which would then prevent the use of Gd to achieve complete cancellation. The magnetisation measurements do not reveal whether the 4$f$ spin or orbital moment dominates, so for that we turn to the present X-ray magnetic circular dichroism (XMCD) investigation.

The conducting characteristics of SmN, unlike SmAl$_2$, SmZn or SmCd, points to it being a semiconductor, as appears to be common among the rare-earth nitrides.~\cite{PhysRevB.73.235335,PhysRevB.75.045114,PhysRevB.76.245120,PhysRevB.84.235120} The implication is that the FM state relies on indirect inter-ion 4$f$ exchange involving the Sm 5$d$ and N 2$p$ orbitals, leading to a lower T$_c$ than in metallic compounds. Clearly FM semiconductors will show distinct advantages for integration with conventional semiconductors, such as the possibility of impedance matching for spin injection.

The very small moment of SmN makes conventional magnetometer studies especially difficult, which as discussed above led to a long-term uncertainty concerning the magnetic order in SmN. The problem is exacerbated in thin film geometries that are of current interest for devices, for there is then very little material available for the measurement. In contrast XMCD measures not the total moment but rather probes the spin- and orbital-moments independently, with atomic-species and orbital-shell selectivity. We thus have investigated SmN using XMCD at the Sm L$_{2,3}$ edges, which concerns a 2$p\rightarrow5d$ promotion of an electron on the Sm$^{3+}$ ion. Work at the M$_{4,5}$ edges would access the 4$f$ moments directly, but the short attenuation length of 1 keV electrons renders the study problematic through the passivating cap that is required to prevent oxidation in the rare-earth nitrides. However, the strong intra-ion 4$f$ - 5$d$ exchange interaction facilitates the L-edge investigation of the 4$f$ spin alignment through the passivating cap. The L-edge results then benefit from comparison with calculated XMCD spectra to extract reliable 4$f$ alignment information. The calculation references the XMCD spectrum to the 4$f$ spin alignment, while the data are referenced to the applied field direction, and this permits for the determination of the spin alignment relative to the applied field.

\section{Experimental and theoretical details}

\subsection{\label{sec:experiment}Sample preparation and experimental setup}

The epitaxial SmN film used in this study was grown by evaporating Sm in the presence of nitrogen gas onto (111) MgO held at a temperature of ${\sim}$400 $^{\circ}$C. The (111)-oriented film was grown at a rate of 100 nm hr$^{-1}$ with a pressure of 10$^{-4}$ mbar of N$_2$; under these conditions the ratio of N$_2$ to Sm flux on the substrate is about 200. After completing the growth, the films were capped \textit{in situ} with AlN. The film growth was monitored by reflection high-energy electron diffraction (RHEED) to confirm the epitaxial nature of the films, and subsequent \textit{ex situ} X-ray diffraction confirmed the orientation and high crystalline quality.
X-ray-absorption spectroscopy (XAS) and XMCD measurements were performed at the Sm L$_{2,3}$ edges on the ID12 beam line at the European Synchrotron Radiation Facility in Grenoble. The signal was detected by total fluorescence yield for which the capping layer is not an impediment at L-edge energies. The film was mounted at close to grazing incidence on the cold finger of a He gas constant flow cryostat, which has been inserted in the bore of a 6~T superconducting split-coil magnet. Temperature was monitored closely adjacent to the sample on the cold finger.

\subsection{\label{sec:computational}Crystal structure and computational details}

The band structure and XMCD calculations have been performed for the face-centered cubic structure of
sodium chloride (space group $Fm\overline{3}m$, No. 225) with lattice constant
$a$=5.1~\AA\, using fully relativistic spin-polarized
linear-muffin-tin-orbital (LMTO) method~\cite{PhysRevB.12.3060,Antonov1995} with the combined
correction term taken into account. We used the Perdew-Wang~\cite{PhysRevB.45.13244}
parametrization for the exchange-correlation potential. Brillouin zone integrations were performed using the improved tetrahedron method~\cite{PhysRevB.49.16223}
and charge self-consistency was obtained with 349 irreducible $\mathbf{k}$
points. To improve the potential we include additional empty spheres. The
basis consisted of Sm $s$, $p$, $d$ and $f$; N $s$, $p$, and $d$; and empty
sphere $s$ and $p$ LMTO's.

The intrinsic broadening mechanisms have been accounted for by folding
XMCD spectra with a Lorentzian. For the finite lifetime of the core
hole a constant width $\Gamma_c$, in general form,~\cite{fuggle1992}
has been used. The finite resolution of the spectrometer
has been accounted for by convolution with a Gaussian of width 0.6 eV.

In order to simplify the comparison of the theoretical x-ray isotropic
absorption spectra of SmN to the experimental ones we take into
account the background intensity which affects the high energy part of
the spectra and is caused by different kinds of inelastic scattering
of the electron promoted to the conduction band above the Fermi level
due to x-ray absorption (scattering on potentials of surrounding
atoms, defects, phonons, etc.). To calculate the background spectra we
used the model proposed by Richtmyer {\it et al.}~\cite{PhysRev.46.843} (for
details see Ref.~\onlinecite{antonov2006}).

We have adopted the LSDA+$U$ method~\cite{PhysRevB.44.943} as a different level of
approximation to treat the electron-electron correlations. We used the
rotationally invariant fully relativistic LSDA+$U$ method.~\cite{PhysRevB.67.155103} The effective on-site
Coulomb repulsion $U$ was considered as an adjustable parameter. We used $U$ =
8 eV. For the exchange integral $J$ the value of 0.66~eV estimated from
constrained LSDA calculations was used.

\section{\label{sec:results}Results}
\begin{figure}[htbp]
\begin{subfigure}{0.49\textwidth}
\includegraphics[width=1.0\textwidth]{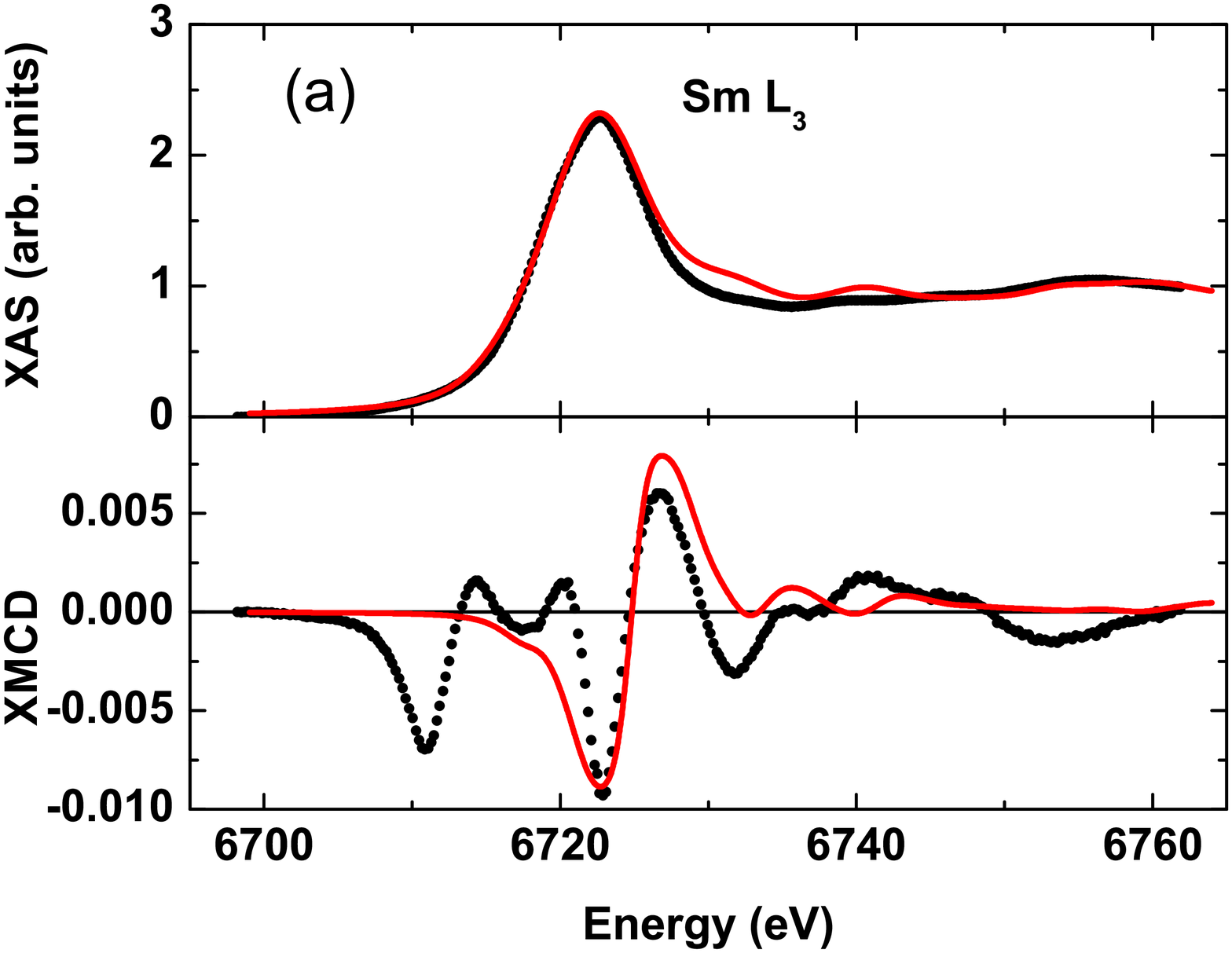}
\label{fig:XASSmL3}
\end{subfigure}
\begin{subfigure}{0.49\textwidth}
\includegraphics[width=1.0\textwidth]{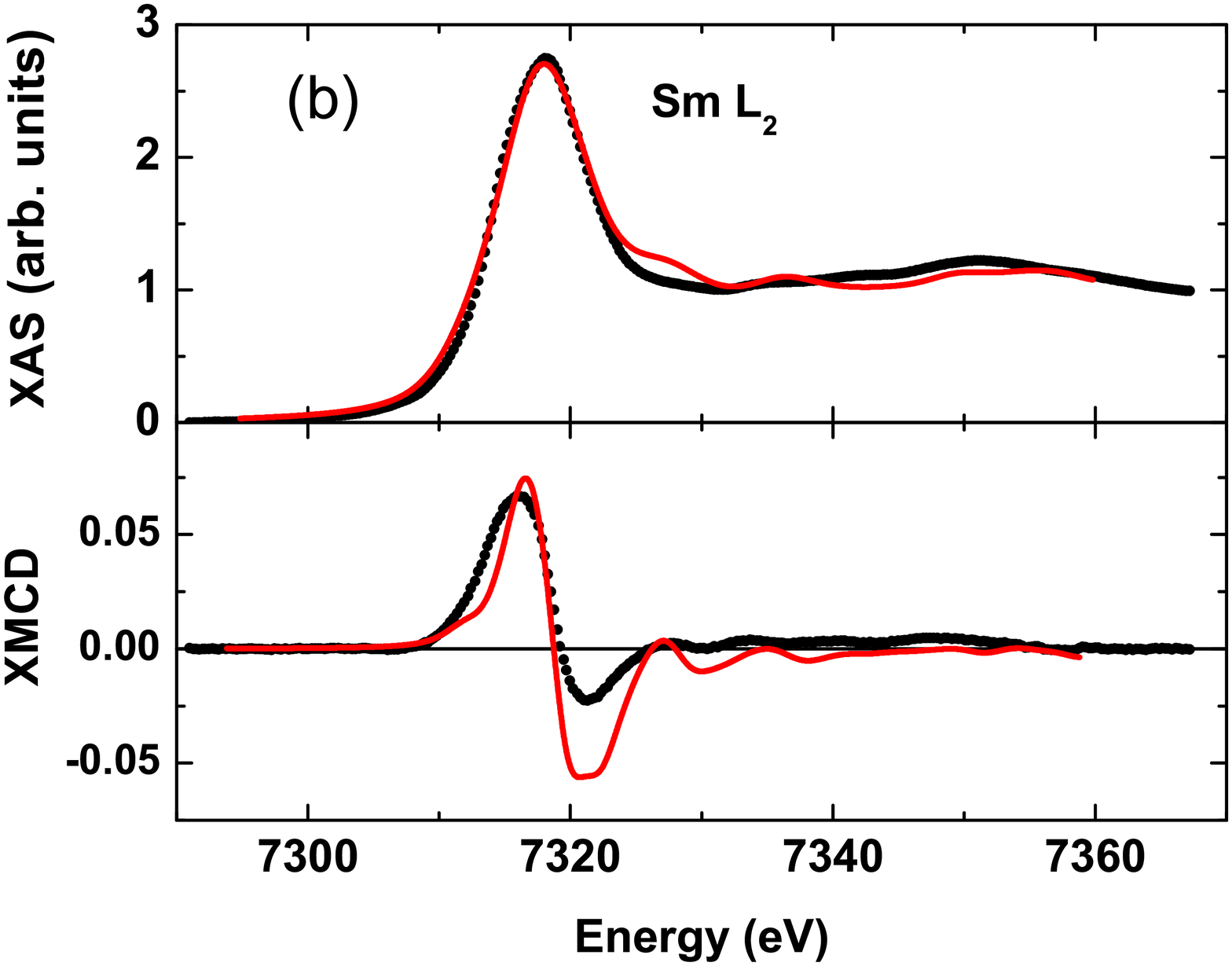}
\label{fig:XASSmL2}
\end{subfigure} 
\caption{(Color online) Measured (symbols) and calculated (lines) XAS and XMCD spectra at (a) the L$_3$ and (b) the L$_2$ edges at 15 K in a field of 6 T. XMCD data are quoted relative to the amplitude of the first prominent (white) peak in XAS. The calculated XMCD spectra have been sign reversed and in the case of the L$_3$ spectrum scaled by a factor 1/3.}
\label{fig:XASandXMCD}
\end{figure}

Sm L$_{2,3}$ XAS and XMCD spectra taken at a temperature of 15~K and in a 6~T magnetic field are compared in {Fig.}~{\ref{fig:XASandXMCD}} with simulated spectra. The computed XAS spectra were adjusted to be in quantitative agreement with the experimental results to achieve an excellent match. The data and calculation both refer to the FM phase, but XAS data taken at higher temperature show no significant difference in the PM phase. Both the computed and experimental XMCD are normalised to the XAS results, so one can make absolute amplitude as well as spectral comparisons. Here the agreement is imperfect, so in {Fig.}~{\ref{fig:XASandXMCD}} the computed XMCD spectra have been scaled by the factors listed in the figure to fit at least the most prominent features in the data. In the case of L$_2$ XMCD, the agreement is adequate if we scale by only -1; i.e., the spectral shape and amplitude are in reasonable agreement, but the sign is incorrect. A similar sign change applies to the L$_3$ edge, though in this case the measured amplitude is a factor of three smaller and there are further features at lower energy; these will be discussed below. The calculation sets the spin axis coincident with that about which the helicity is defined, but in contrast the experimental convention is to set the applied field antiparallel to the x-ray propagation direction.~\cite{Parlebas2006} The two definitions coincide when the spin moments are aligned directly by the field, so the sign difference here immediately identifies the spin direction as lying along the field; the spin magnetic moment is aligned in opposition to the field. The argument is not new; note that the sign of Sm XMCD has already been used to investigate the alignment in SmAl$_2$,~\cite{Parlebas2006,PhysRevB.82.174421} and furthermore it is in agreement with data on GdN, where there is no doubt that the spin moment aligns with the field.~\cite{PhysRevB.73.214430} The net magnetic moment in SmN lies along the 4$f$ orbital magnetic moment.

Aside from a sign reversal, it is notable that the XMCD signal is larger by nearly a factor of ten at L$_2$ than at L$_3$. Such an unbalanced branching ratio is not uncommon for the light rare earths, and results from a dependence of the 5$d$ orbital cloud on the 4$f$-5$d$ exchange; the majority spin orbital is slightly more localised toward the nucleus than that of the minority spin, and hence it overlaps the $2p$ orbital more strongly.~\cite{Matsuyama1997,PhysRevLett.78.1162,Parlebas2006} That is also reflected in the LSDA+$U$ simulation of {Fig.}~{\ref{fig:XASandXMCD}}, although the L$_2$:L$_3$ ratio is underestimated to be ${\sim} 3$. Furthermore there is a strong electric quadrupole (2$p\rightarrow4f$) signal extending to some 12 eV lower in energy. In part the prominence of the quadrupole signal at L$_3$ follows the weakening of the 2$p\rightarrow5d$ electric dipole signal in the light rare earths.~\cite{Parlebas2006} The lowest energy 4$f$ states within the band structure of SmN do not lie so far as 12 eV below the 5$d$ conduction band,~\cite{PhysRevB.75.045114} but the 4$f$ XAS/XMCD features are shifted lower in energy because the 4$f$ electrons interact more strongly with the core hole.~\cite{Parlebas2006} It is notable that the quadrupole signal follows the dipole signal closely as the temperature is raised; the polarisation of the 5$d$ shell is as expected determined by intra-atomic 4$f$-5$d$ exchange.

\begin{figure}[htbp]
\begin{subfigure}{0.49\textwidth}
\includegraphics[width=1.0\textwidth]{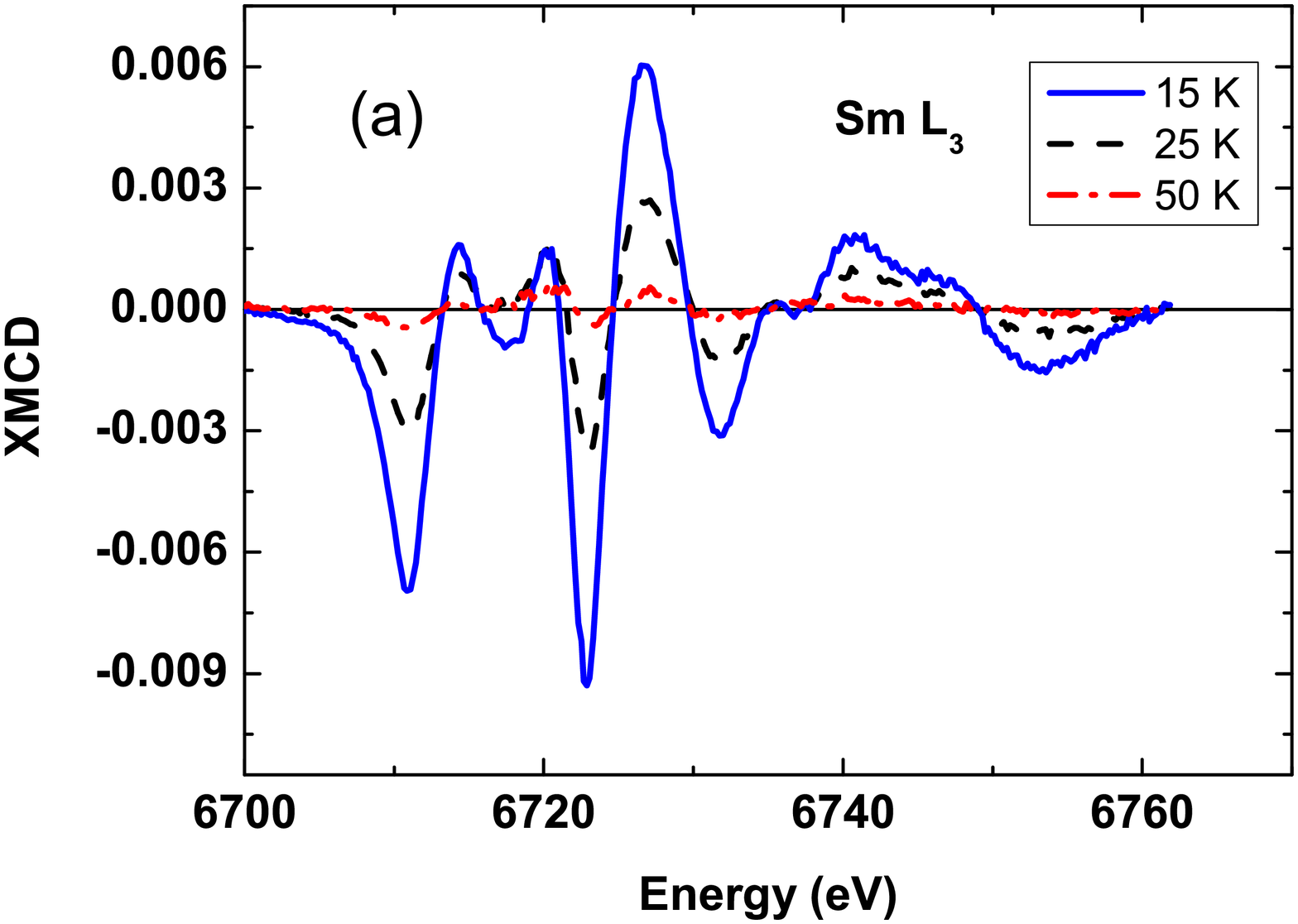}
\label{fig:XMCDSmL3vsT}
\end{subfigure} 
\begin{subfigure}{0.49\textwidth}
\includegraphics[width=1.0\textwidth]{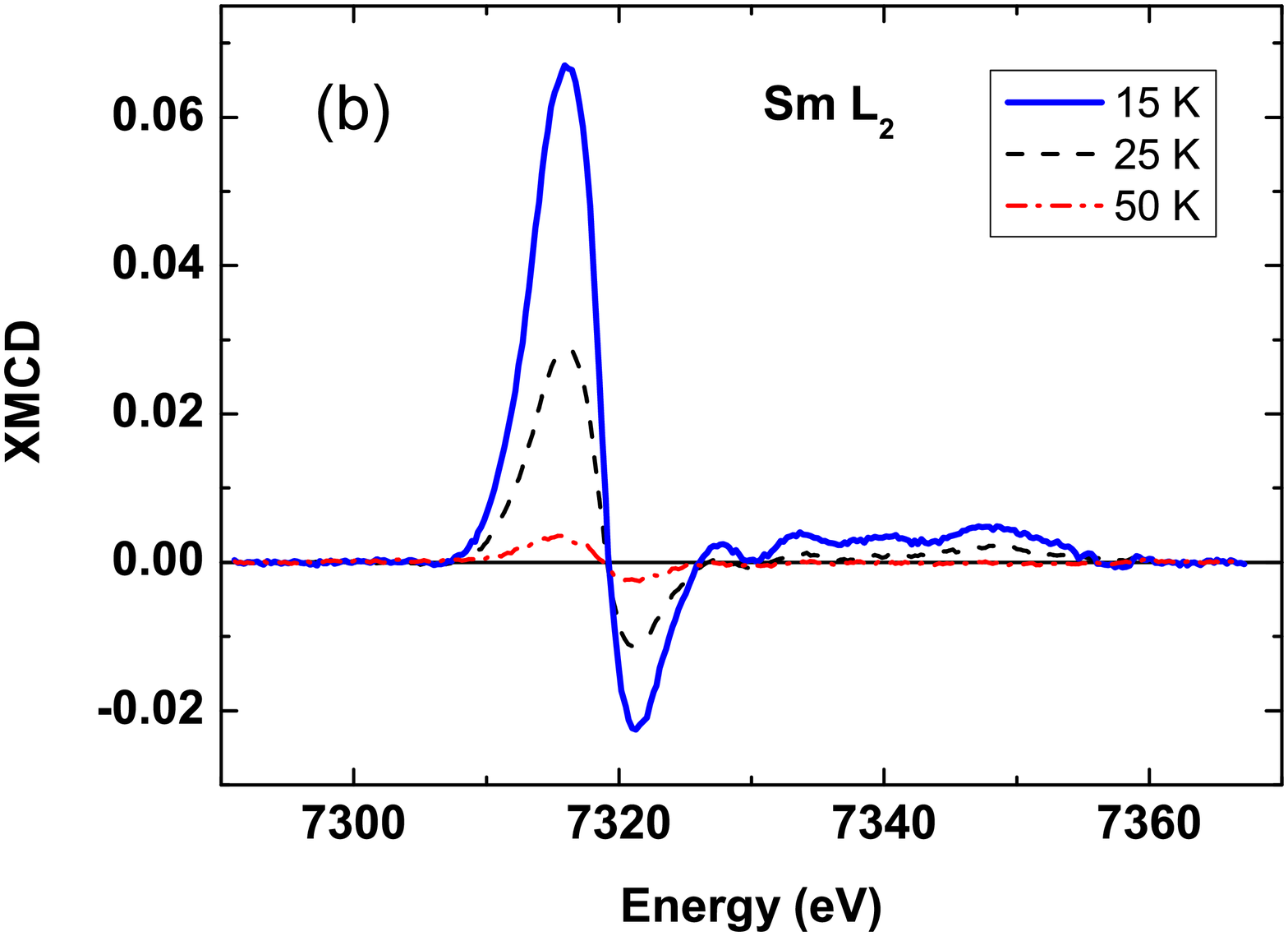}
\label{fig:XMCDSmL2vsT}
\end{subfigure} 
\caption{(Color online) Temperature dependence of (a) the L$_3$ and (b) the L$_2$ edge XMCD spectra in the ferromagnetic (15, 25 K) and paramagnetic (50 K) phases. The magnetic field was 6 T.}
\label{fig:XMCDvsT}
\end{figure}

\begin{figure}[htbp]
\includegraphics[width=0.8\textwidth]{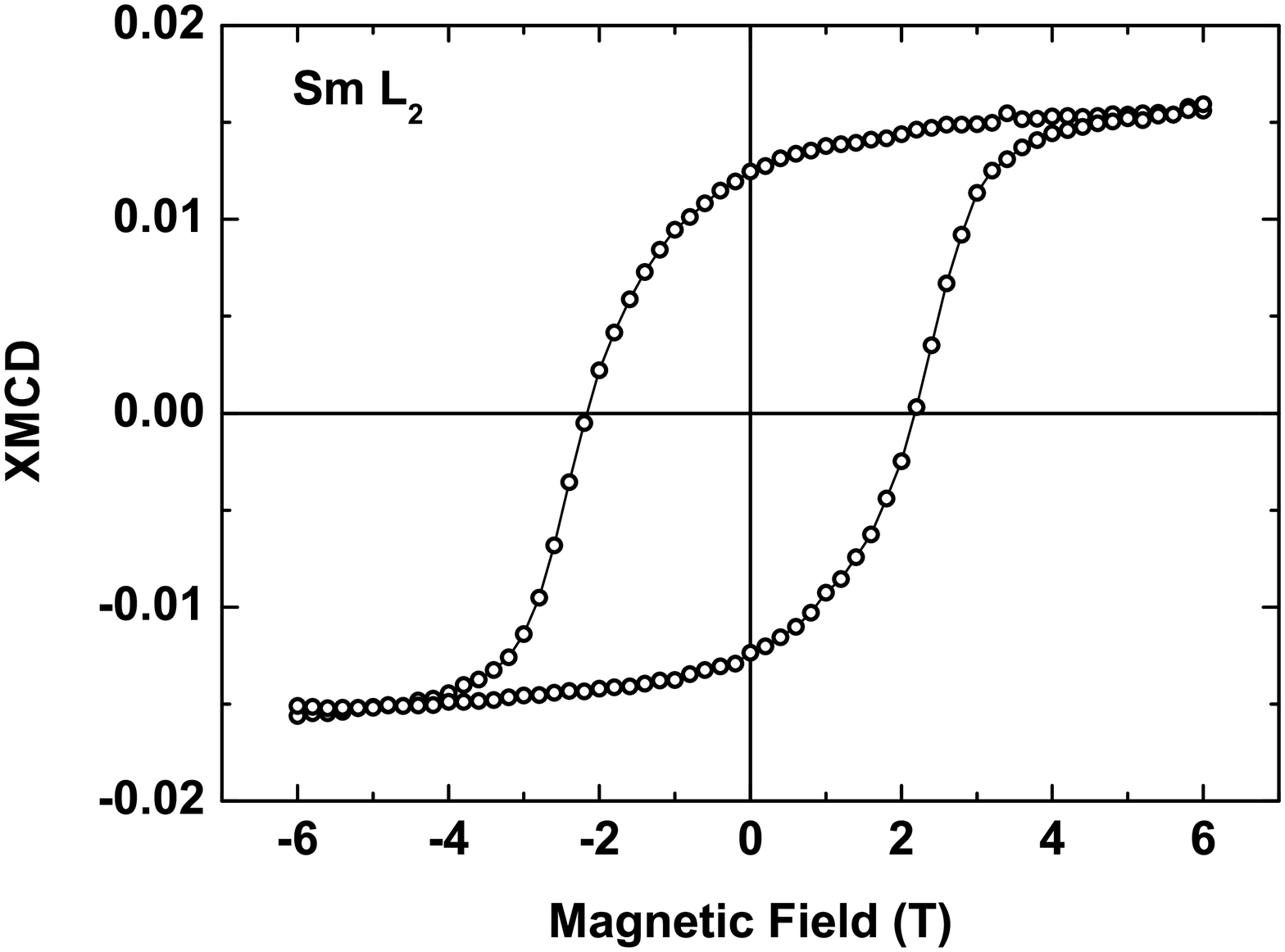}
\caption{Field dependence of L$_2$ edge XMCD amplitude at 15 K showing wide hysteresis loop.}
\label{fig:MHSm}
\end{figure}

In {Fig.}~{\ref{fig:XMCDvsT}} is a comparison of the L$_2$ and L$_3$ edge spectra collected in the FM phase at 15 and 25~K, and at 50 K in the PM phase. The spectral shape and sign is unchanged across that temperature range, implying that the net moment lies along the 4$f$ orbital and antiparallel to the 4$f$ spin moment in the PM as well as the FM phase.  The data show a reduction by a factor of 2 between 15 and 25 K, in agreement with the 27 K T$_C$ in SmN. The signal dies away further in the PM phase at 50 K. The hysteresis at 15 K ({Fig.}~{\ref{fig:MHSm}}) confirms the magnetisation measurements,~\cite{PhysRevB.78.174406} with a coercive field of $>$2~T signalling the weak coupling of the small net moment to an external field.

\begin{figure}[htbp]
\includegraphics[width=0.8\textwidth]{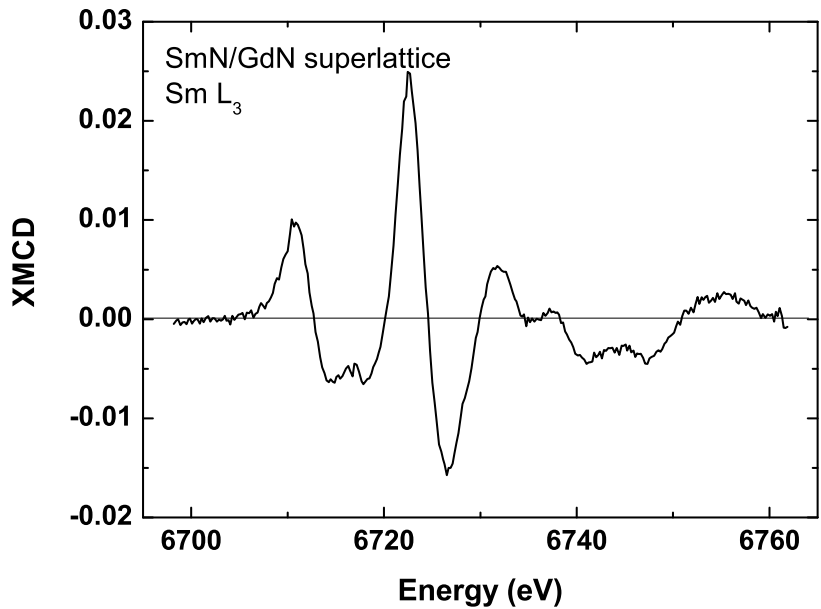}
\caption{SmL$_3$ edge XMCD in 1.5~nm SmN layers embedded in GdN at 15 K.}
\label{fig:XMCDSL}
\end{figure}

To confirm that the sign of the XMCD spectra above signal that the Sm$^{3+}$ spin lies antiparallel to the field, we have investigated also very thin SmN layers embedded in GdN, an epitaxial superlattice of 12$\times$(1.5~nm SmN/10~nm GdN). The 7 $\mu_B$ spin on Gd$^{3+}$ ensures that its alignment is along the applied field. The inter-ion exchange that precipitates FM behaviour can then be expected to align the Sm and Gd spin moments, thus reversing the sign of the Sm XMCD signatures. Indeed that is exactly what is seen in {Fig.}~{\ref{fig:XMCDSL}}, showing the Sm L$_3$ signal with the sign reversed relative to homogeneous SmN. The dipole (2$p\rightarrow5d$) amplitude is also a factor of ${\sim}2.5$ larger in the superlattice; clearly the 5$d$ spin is strongly aligned by inter-ion exchange across the SmN/GdN interface. The quadrupole (2$p\rightarrow4f$) signal appears marginally stronger, but by only 30~\%. A comparison with the XMCD at the Sm L$_2$ edge is prevented in this case by masking from magnetic EXAFS (extended X-ray absorption fine structure) above the Gd L$_3$ edge.

\clearpage

\section{\label{sec:conclusions}Conclusions}

We have reported XMCD in SmN, focusing on an investigation of the Sm 5$d$ spin alignment at the L$_{2,3}$ edges. The data are interpreted with the aid of an LSDA+$U$ calculation of the L$_{2,3}$ edge XMCD. Except for the sign the calculated and measured XMCD spectra are in excellent agreement at the L$_2$, but they differ by a factor of three at the L$_3$ edge, related to the unusual branching ratio for these transitions. There is also a prominent quadrupolar feature at the L$_3$ edge, with a visibility that is enhanced by the relatively weak dipolar spectrum at that edge. 
SmN is known to possess a strikingly small FM moment of about 0.03~$\mu_B$ per Sm ion; the 4$f$ spin and orbital magnetic moments cancel to within less than 1~\% of their individual magnitudes. The sign of the measured XMCD spectra are opposite to that calculated, identifying that the residual magnetic moment in SmN is directed anti-parallel to the 4f spin magnetic moment. The material is in that respect similar to SmAl$_2$, and it is likely that the moment could be reduced to zero by substituting about 0.5~\% Gd in the cation site. 
Temperature-dependent XMCD is in excellent agreement with the 27 K T$_c$ that has been determined by magnetisation studies. The field dependence shows the same large coercive field determined by magnetisation studies, and as is expected under the very weak coupling of an applied field with the vanishingly small residual moment.

\begin{acknowledgments}
We acknowledge financial support from the NZ FRST
(Grant No. VICX0808) and the Marsden Fund (Grant No.
08-VUW-030). The MacDiarmid Institute is supported by the New Zealand
Centres of Research Excellence Fund. E.A. thanks the Alexander-von-Humboldt foundation for support through a fellowship.
\end{acknowledgments}

%\bibliography{JabRef/Literatur}

%

\end{document}